\documentclass{pic2012}
\def\runauthor{"Blahoslav Pastir\v{c}\'{a}k et al"}
\def\shorttitle{JEM-EUSO experiment for EECR observation}
\begin{document}

\title{JEM-EUSO experiment for extreme energy cosmic ray observation}

\author{Blahoslav Pastir\v{c}\'{a}k, Pavol Bob\'{i}k and Karel Kudela \\
for the JEM-EUSO collaboration}

\address{Institute of Experimental Physics SAS\\
Watsonova 47, 040 01 Ko\v{s}ice, Slovak Republic\\
E-mail: slavo@saske.sk }

\maketitle

\abstracts{
    The planned JEM-EUSO (Extreme Universe Space Observatory onboard the ISS Japanese Experimental Module) will measure the energy spectra of cosmic rays 
    up to the range of $10^{21} ~eV$ and will search for direction to their sources. It will observe the extensive air showers generated in the atmosphere 
    by high energy cosmic ray primary particle from the space. The instantaneous aperture of the telescope  will exceed by one order the aperture of the largest 
    ground based detectors. JEM-EUSO apparatus is a large telescope with a diameter of $2.5 ~m$ with fast  UV camera. Slovakia is responsible for the determination 
    of the UV background, which influences the operational efficiency of the experiment and for the analysis of  fake trigger events. 
    }

\section{Extreme Universe Research}

    The origin and existence of the extreme energy cosmic rays (EECRs) and the physical mechanism of their acceleration to very high energies are an open puzzle
    in contemporary astroparticle physics. The highest observed cosmics rays energy is about $3 \times 10^{20} ~eV$, which exceeds by 8 orders the CERN LHC energy scale. 
    
    Such extremely powerful sources capable of accelerating cosmic rays up to EECRs energies have to be confined within a limited range of distances set by the 
    Greisen-Zatseptin-Kuzmin cuttof of $6 \times 10^{19} ~eV)$ \cite{gzk}, which is caused by interactions of cosmic rays with the cosmic microwave background. Possible 
    indications of sources or excesses in arrival direction distribution of EECRs have been claimed by several ground based experiments \cite{auger}, \cite{agasa}, 
    \cite{ta}, \cite{hires} implying that the sources have to be up to several tens $MPc$ far. Possible EECRs sources are supernovae, pulsars, gamma ray bursts, active 
    galactic nuclei and recent collisions of radiogalaxies. However, most of these candidates are incapable of accelerating particles beyond $10^{20} ~eV$ by any 
    known acceleration mechanism.
    
\begin{figure}[ht]
  \begin{minipage}[t]{6cm}
  \includegraphics[width=6cm]{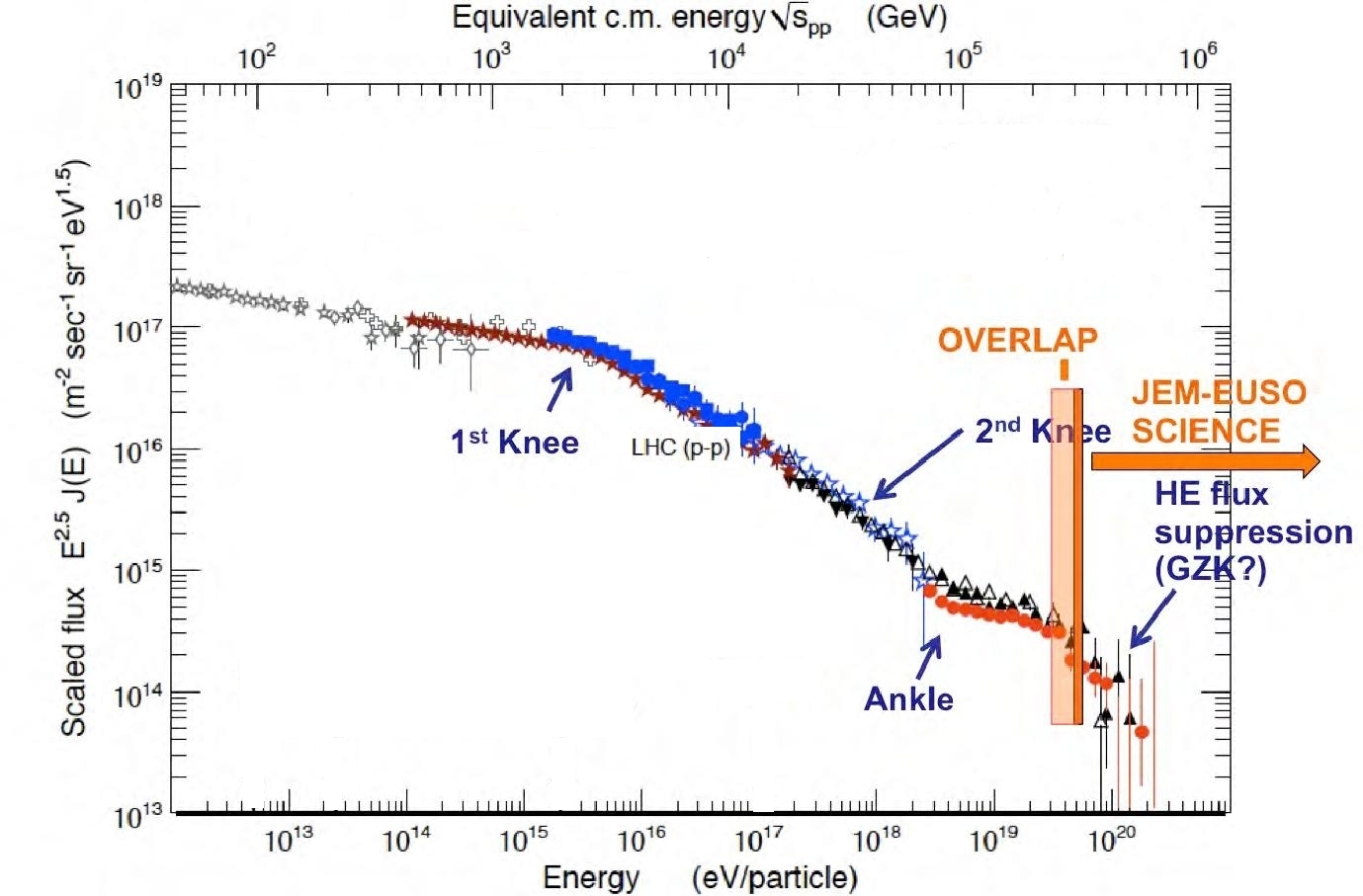}
  \caption{Scaled energy spectrum of cosmic rays}
  \label{Fig:Flux}
  \end{minipage}
  \qquad
  \begin{minipage}[t]{5cm}
  \includegraphics[width=5cm]{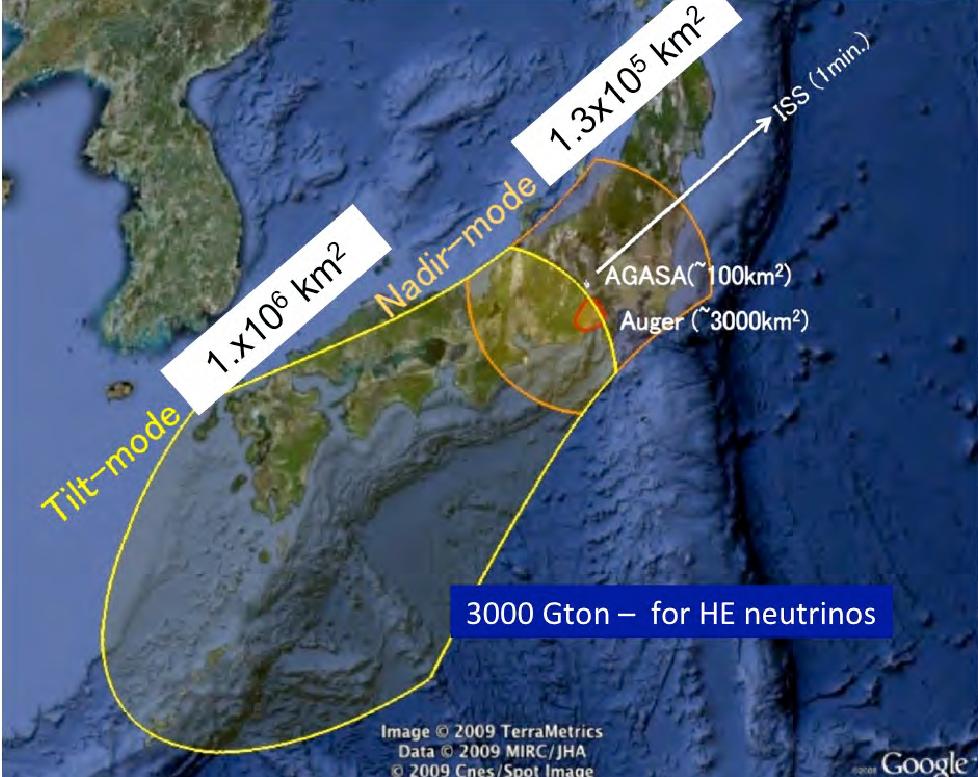}
  \caption{The instantaneous aperture observed by the JEM-EUSO telescope compared with some ground based experiments}
  \label{Fig:Aperture}
  \end{minipage}
  \end{figure}

    To identify the sources of EECRs, the measurements of the energy spectrum and arrival directions of such particles are needed. Although low energy charged 
    particles are bent by magnetic field in intergalactic and galactic space so that the directional information of their origin is lost, the highest energy particles 
    are barely bent, thus retaining the information of the direction to the origin. The EECRs flux is  exceptionally low, of the order 
    of $\it 1~ particle /km^{2} /sr /century$ at energies  above Greisen-Zatsepin-Kuzmin energy. At the high end of the spectrum, $E > 10^{20} eV$, it 
    reduces even to  about $\it 1~ particle /km^{2} /sr /millenium$ (Figure \ref{Fig:Flux}). This challenging extreme energy region is the scope of the Extreme 
    Universe  Space Observatory (EUSO) attached to the Japanese Experiment Module (JEM) on board the International Space Station (ISS) \cite{B}.

    The size of the observational area is a critical factor for detecting the rare EECRs events. 
    Currently, the leading observatories of EECRs are ground based, that cover large areas with particle detectors overlooked by fluorescence 
    telescopes. The largest one is the Pierre Auger Observatory in Argentina with a surface detector array covering $3000~ km^{2}$ which accumulates anually 
    about $ 6 \times 10^{3}~ km^{2}~ sr ~yr$ of exposure \cite{auger1}. The more recently constructed Telescope Array (TA) covers $700~ km^{2}$ which should accumulate
    annually $1.4 \times  10^{3}~ km^{2} ~ sr ~ yr$ of exposure. Although extremely large, ground based observatories have nearly reached the maximum extent possible 
    on earth. Space observatory makes a giant leap in the area size observed (Figure \ref{Fig:Aperture}). JEM-EUSO mission explores the origin EECRs and explores the limits 
    of the fundamental  physics, through the observations of their arrival directions and  energies. An additional relevant advantage comes from the orbit, as JEM-EUSO will monitor with 
    a rather uniform exposure both hemispheres thus minimizing the systematic uncertainties that strongly affect any comparison between different observatories exploring, 
    from ground, different hemispheres.

\section{JEM-EUSO Experiment}

    The proposal of the JEM-EUSO mission  \cite{A}, \cite{B}, \cite{C}, \cite{D}  is an common project between the space  agencies  JAXA, NASA, ESA, Roskosmos and 13 collaborating countries (77 institutions, 
    over 250 researchers). The leading  country is  Japan, which provides  the basic infrastructure including a vehicle HII-B, a spaceship HTV and the position 
    for detector emplacement onboard the  ISS  Japanese Experimental Module Kibo.  It will operate  minimum 3 years with starting date of 2017.
    
    This innovative space mission will use the Earth`s atmosphere as a detector of cosmic ray showers. JEM-EUSO exploits the fluorescence light that is 
    emitted during the development of the Extensive Air Shower (EAS) initiated by a primary cosmic ray particle in the atmosphere to estimate the energy 
    and arrival direction of EECRs. JEM-EUSO will measure the energy spectra of cosmic rays up to $10^{21} eV$ and will search for direction to their sources. 
    By observing from space the fluorescence and Cherenkov light emitted by EAS, the species, energy and direction of the primary is well determined.

\begin{figure}[htb]
  \begin{minipage}[t]{5cm}
  \includegraphics[width=5cm]{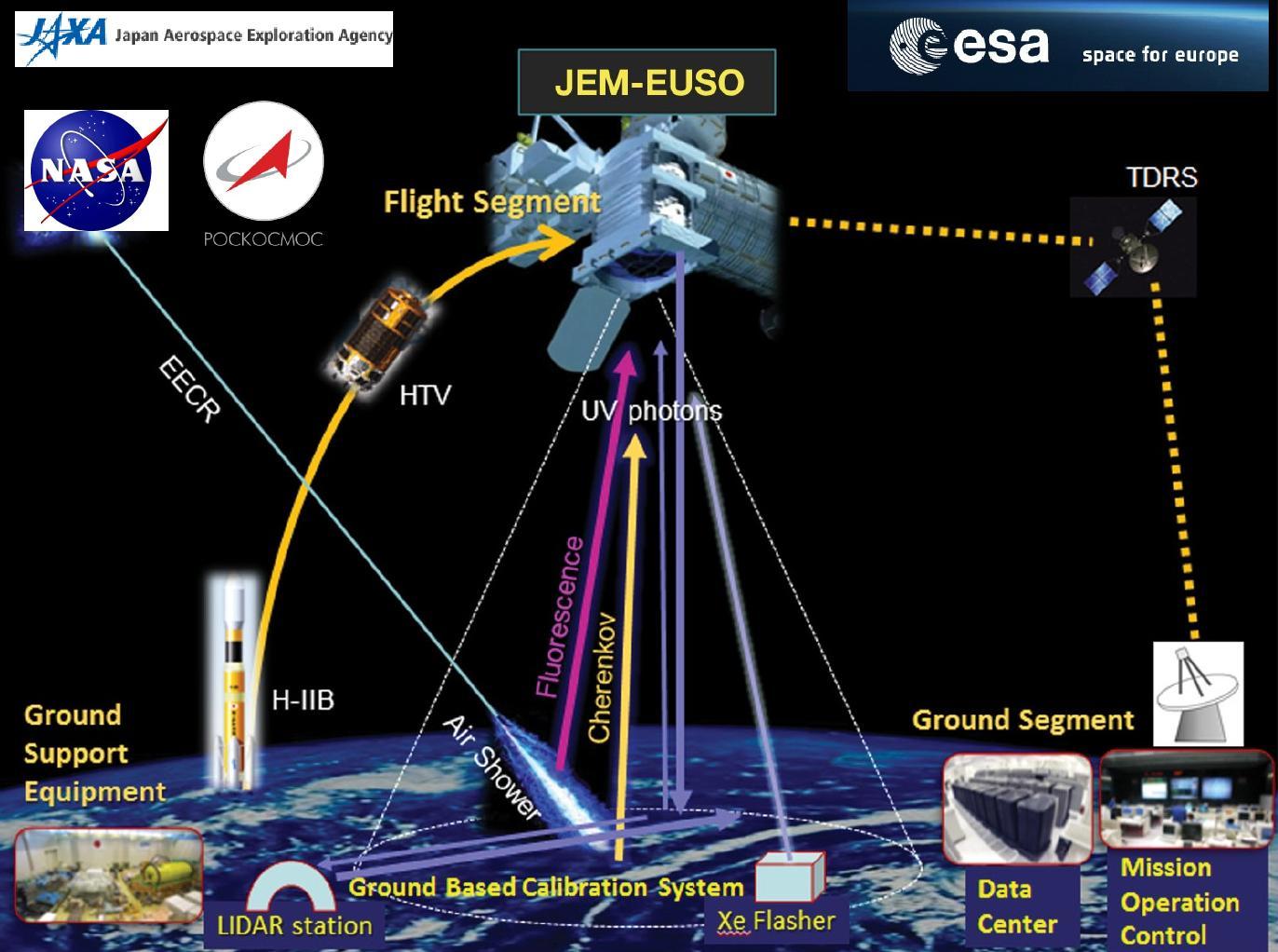}
  \caption{JEM-EUSO overview and principle of operation}
  \label{Fig:Principle}
  \end{minipage}
  \qquad
  \begin{minipage}[t]{6cm}
  \includegraphics[width=6cm]{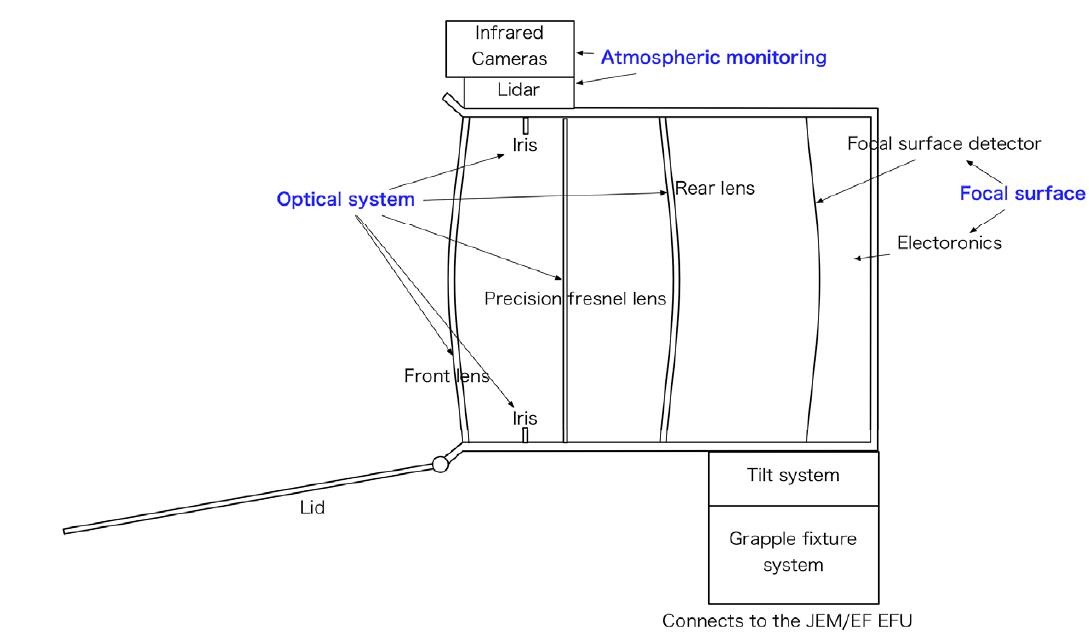}
  \caption{Telescope}
  \label{Fig:Telescope}
  \end{minipage}
  \end{figure}

  Figure \ref{Fig:Principle} illustrates the EECRs observation principle in the JEM-EUSO mission. On the orbit of altitude $400~km$   
  the JEM-EUSO telescope detects fluorescence and Cherenkov light from the EAS. The former directly heads to the telescope. The latter 
  is observed either because of scattering in the atmosphere or because of diffuse reflection from the surface or the cloud on which 
  the Cherenkov beam impacts. An incoming $10^{20}~ eV$ EECR produces an order of $10^{11}$ particles at shower maximum during the EAS development. 
  The secondary particles are still relativistic and the charged particles, most dominantly electrons, excite the nitrogen molecules to emit UV fluorescence 
  light of characteristic lines in the wavelength range of $300~ - ~430~ nm$. The total yield of $4~ - 5~ photons/ m/ electron$ \cite{yield1}, \cite{yield2} 
  is almost independent of the altitude. Along the development of a $10^{20}~ eV$ EECR shower order of $10^{15}$ photons are isotropically emitted. Seen from $∼~ 400~ km$ 
  distance, the solid angle of the telescope with a few $m^{2}$  aperture is about  $10^{−11}$ sr.  In the clear atmospheric condition, therefore, 
  it results in few thousands photons reaching the pupil of the telescope. JEM-EUSO is designed to record not only the number of photons but also their direction 
   and arrival time. 
  
    Technically JEM-EUSO is a large telescope with a diameter 2.5 m with fast UV camera. The telescope \cite{telescope} is an extremely fast (camera takes 
    400000 frames/s), highly pixelized ( more than 300000 pixels) instrument. These allow to record cascade in angle and time. Camera will record “video clips” 
    of fast moving UV tracks sensing the fluorescence light representing the temporal development of EAS.
    
   The telescope (Figure \ref{Fig:Telescope}) mainly consists of four parts: the photon collecting optics \cite{optics}, the focal 
   surface (FS) detector \cite{FS}, the electronics \cite{electronics}  and the mechanical structure \cite{MS}. 
   Since the development of the EAS and the intensity of the observed light depend on the transmittance of the atmosphere, the cloud coverage and the height 
   of cloud-top, JEM-EUSO is equipped with an atmospheric monitoring (AM) system \cite{AM}. To estimate as precisely as possible the atmospheric conditions 
   and the effective observation aperture with high accuracy, AM system will consist of an infrared (IR) camera, a LIDAR system and may benefit the use 
   of UV data acquired continuously by the telescope itself.

\section{Our Contribution}

   In Slovakia the Institute of Experimental Physics is participating in JEM-EUSO experiment preparation. We are responsible for several tasks.
   The main are:
   \begin{itemize}
    \item The estimation  of the UV background on the night side of the Earth Sources of the backgound 
    
    \item The determination of the JEM-EUSO operational efficiency fraction of time when  monitoring UV compared to full time on orbit. Above 
       mentioned UV background sources together with ISS operation schedule had to be taken into account in the model of JEM-EUSO operational 
       efficiency
    
    \item The fake trigger event simulations and analysis  
   \end{itemize}
   
   The main sources of the UV background are the reflections from sky (Moon, stars, planets), man made lights, lightnings, airglow, aurora, meteorites.
   Most of them are yet included into the simulations.
   
   The operational efficiency is a fraction of the time when monitoring UV compared to full time on orbit.
   Above discussed contributions to the UV background together with ISS operation schedule had to be taken into account in the model of JEM-EUSO operational efficiency.
   
    The most relevant background component affecting the observational duty cycle is the effect of the Moon. We estimated the moonlight component from
    the phase of the Moon and its apparent position as seen from the ISS. In our approach the ISS trajectory provided by NASA SSCweb was traced with
    1-min time steps and the moonlight at the top of the atmosphere was estimated according to \cite{bg1}. For every position of the ISS in the period from 2005
    till 2007 the zenith angle of the Sun, and that of the Moon as well as the moon phase angle were calculated. The contribution to background level due to reflected 
    moonlight was evaluated in the way described in the following, modified from \cite{bg2}.
    
\begin{figure}[!thb]
    \begin{center}
    \mbox{\epsfig{file=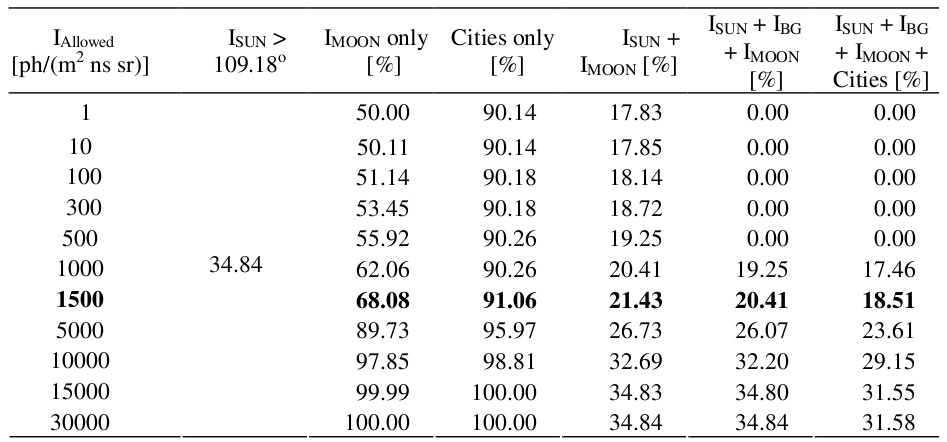,width=.75\linewidth}}
    \caption{Summary of all calculated effects contributing  to operational efficiency }
    \label{DC}
    \end{center}
    \end{figure}

    The summary of all effects taken into account in evaluation of operational efficiency up to now is presented in Figure \ref{DC}.

\begin{figure}[!thb]
    \begin{center}
    \mbox{\epsfig{file=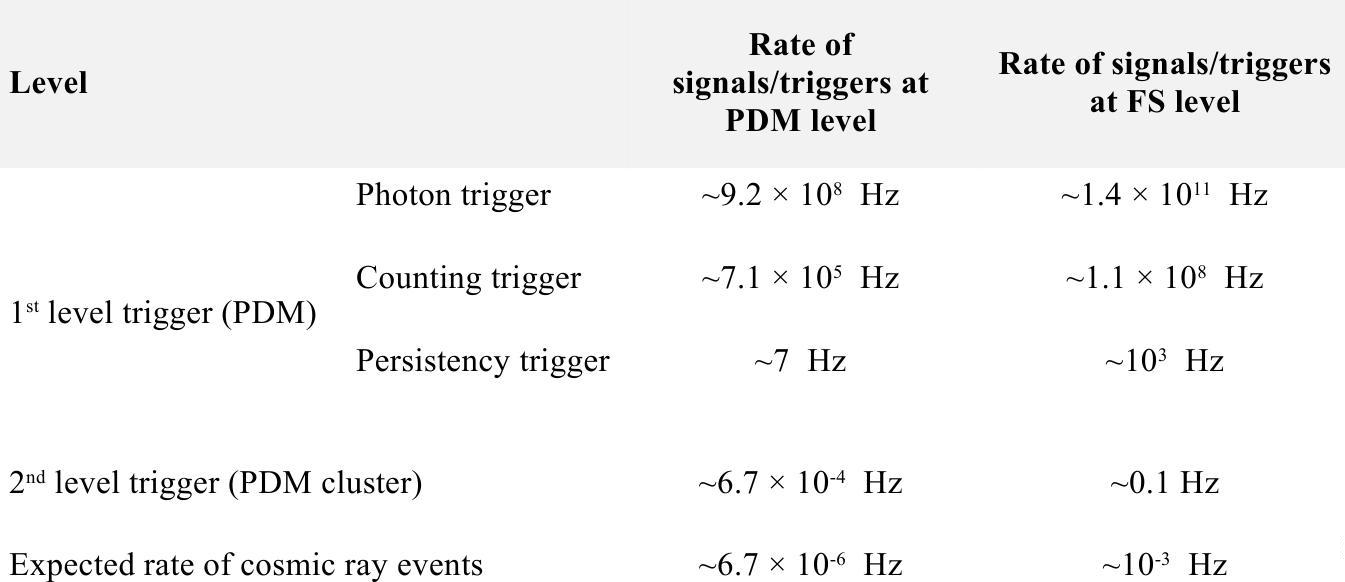,width=.75\linewidth}}
    \caption{The description of the rate reduction on different trigger levels}
    \label{Trig1}
    \end{center}
    \end{figure}

       The last issue among our responsibilities in JEM-EUSO experiment preparation is the simulation and following analysis 
       of the fake trigger events.
       The goal of the trigger system is to detect the occurrence of scientifically valuable signal from the EECRs events 
       among very huge background noise detected by JEM-EUSO \cite{trig1}. Its expected rate is up to $10^{11} Hz$ , see Figure \ref{Trig1}.
       The UV background registered by JEM-EUSO is randomly distributed with Poissonian character. We study, if these random processes produce fake pattern, 
       which could be mistakenly interpreted  as EECRs events. We are simulating huge amount of measurements on one model PDM (Photo Detector Module) 
       with only detector noise. In the simulation code two main trigger levels are implemented  and consequently such filtered output is going 
       to be analysed \cite{trig2}. To distinguish between such simulated  fake events and real ECCRSs events and find the probabillity of registration fake event 
       we are applying and developing  pattern recognition methods, especially Hough transform method \cite{trig3}. 

\section*{Acknowledgements}
    The authors acknowledge the Slovak Academy of Sciences (SAS) for the grant 
    supporting the SAS participation in JEM-EUSO as a project related to ESA.



\begin{thebibliography}{0}

\bibitem{gzk}
K.~Greisen Phys.Lett. 16, 748 (1966) \\
G.~T.~ Zatsepin and V.~A.~ Kuzmin, JETP Phys. Lett. 16, 148 (1966)

\bibitem{auger}
J.~ Abraham et al. [The Pierre Auger Collaboration], Phys.Rev.Lett {\bf 101},  101101 arXiv:astro-ph/0703099 (2008) \\
L.Wiencke et al., Proceedings of the 32nd International Cosmic ray Conference, Beijing, arXiv:1107.4806 (ICRC 2011)

\bibitem{agasa}
http://www.akeno.icrr.u-tokyo.ac.jp/AGASA

\bibitem{ta}
Y.~Tsuneda et al. [Telescope Array Collaboration],   Proceedings of the 32nd International Cosmic ray Conference, Beijing, arXiv:1111.2507 (ICRC 2011) \\
http://www.telescopearray.org

\bibitem{hires}
R.U.Abbasi et al. [HiRes Collaboration], Phys.Rev.Lett.100:101101,2008 \\
http://www.cosmic-ray.org

\bibitem{B}
Y.~Takahashi et al., New Journal of Physics, 11, 065009 (2009)

\bibitem{auger1}
J.~ Abraham et al. [The Pierre Auger Collaboration], Nucl. Instr. Meth. A{\bf 613}, 29  (2010)

\bibitem{A}
The JEM-EUSO Mission: Status and Prospects in 2011, arXiv:1204.5065v1 (2011)

\bibitem{C}
T.~Ebisuzaki et al., Nucl.Phys.B, 175-176,237 (2008)

\bibitem{D}
T.~Ebisuzaki et al,Tours Symposium on Nuclear Physics and Astrophysics - VII, pp369-376 (2009)

\bibitem{E} 
F.~Kajino et al., Nuclear Instruments and Methods in Physics Research A63, 422-424 (2010)

\bibitem{yield1}
E.~F.~Arqueros,  J.~R.~Horandel, B.~Keilhauer, Nucl. Instr. Meth. A{\bf 597}, 1  (2009)

\bibitem{yield2}
G.~Lefeuvre et al., Nucl. Instr. Meth. A{\bf 578}, 78  (2007)

\bibitem{telescope}
F.~Kajino et al., Nucl. Instr. Meth. A{\bf 623} 422 (2010)

\bibitem{optics}
A.~Zuccaro Marchi et al., Proceedings of the 32nd International Cosmic ray Conference, Beijing, 852 arXiv:1204.5065 (ICRC 2011)

\bibitem{FS}
Y.~Kawasaki  et al., Proceedings of the 32nd International Cosmic ray Conference, Beijing, 472 arXiv:1204.5065 (ICRC 2011)

\bibitem{electronics} 
M.~Casolino et al.,  Proceedings of the 32nd International Cosmic ray Conference, Beijing, 1219 arXiv:1204.5065 (ICRC 2011)

\bibitem{MS}
M.~Ricci et al., Proceedings of the 32nd International Cosmic ray Conference, Beijing, 0335 arXiv:1204.5065 (ICRC 2011)

\bibitem{AM}
A.~Neronov et al., Proceedings of the 32nd International Cosmic ray Conference, Beijing, 0301 arXiv:1204.5065 (ICRC 2011)

\bibitem{bg1}
P. Bobik et al., Proceedings of the 32nd International Cosmic ray Conference, Beijing,  0886, arXiv:1204.5065. (ICRC 2011)

\bibitem{bg2}
F. Montanet, EUSO-SIM-REP-009-1.2 2004.

\bibitem{trig1}
J. H. Adams Jr., L. A. Anchordoqui, M. Bertaina, et al. Summary Report of JEM-EUSO Workshop at KICP in Chicago. \\
Cite as: arXiv:1203.3451v1 [astro-ph.IM]

\bibitem{trig2}
http://space.saske.sk/JESM/presentations/BP.pdf

\bibitem{trig3}
Hough~P.V.C., In: Instrumentation for High Energy Pysics, p242 (1961)

\end{thebibliography}
\end{document}